\documentclass[12pt,english,twoside]{article}

\usepackage{lastpage}  

\usepackage[T1]{fontenc}
\usepackage[utf8]{inputenc}
\usepackage[letterpaper, margin=2cm]{geometry}
\usepackage{color}
\usepackage{graphicx}
\usepackage{subfigure}
\graphicspath{{graphics_fold/}}
\usepackage{epstopdf} 
\usepackage{times}
\usepackage{etoolbox}
\usepackage{amsmath} 
\usepackage{amssymb}  
\usepackage{fancyhdr}
\usepackage{psfrag}
\usepackage{enumitem}
\usepackage{indentfirst}       
\usepackage{titlesec}          
\usepackage[hidelinks]{hyperref}
\usepackage{babel}
\usepackage{listings}
\usepackage{booktabs}

\addto\captionsenglish{}

\usepackage[colorinlistoftodos]{todonotes} 

\setlist{nolistsep}

\makeatletter
\renewcommand*{\@biblabel}[1]{\hfill#1.}
\makeatother

\makeatletter
\patchcmd{\headrule}{\hrule}{\hrule}{}{}
\patchcmd{\footrule}{\hrule}{\hrule}{}{}
\makeatother

\fancypagestyle{firstpage}{
\fancyhead{}
\fancyfoot{}

}

\makeatletter
\def\maketitle{
  \thispagestyle{firstpage}
  \vspace*{-25mm}  

  \noindent
  \hspace*{-0.8cm} 
  \begin{minipage}[t]{0.55\textwidth}
    \vspace*{3mm}
    \includegraphics[width=5.3cm]{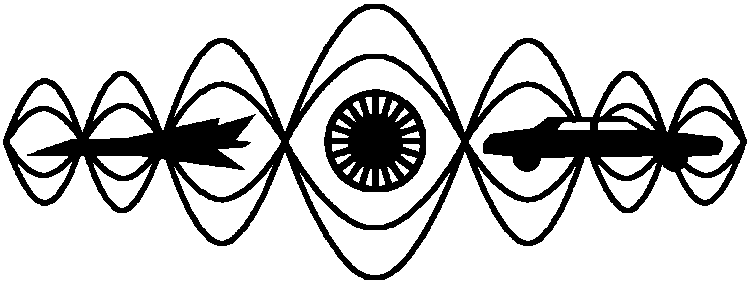}\\[-1mm]
    {\fontsize{10.5pt}{12pt}\selectfont\rmfamily
      \textcolor[rgb]{0.8, 0.0, 0.0}{\textbf{31\textsuperscript{st} International Congress}}\,%
      \textcolor[rgb]{0.45, 0.45, 0.45}{on}\,%
      \textcolor[rgb]{0.0, 0.2, 0.4}{\textbf{Sound and Vibration}}%
    }
  \end{minipage}%
  \hfill
  \begin{minipage}[t]{0.4\textwidth}
    \vspace*{-3mm}
    \begin{flushright}
      \includegraphics[width=5.72cm]{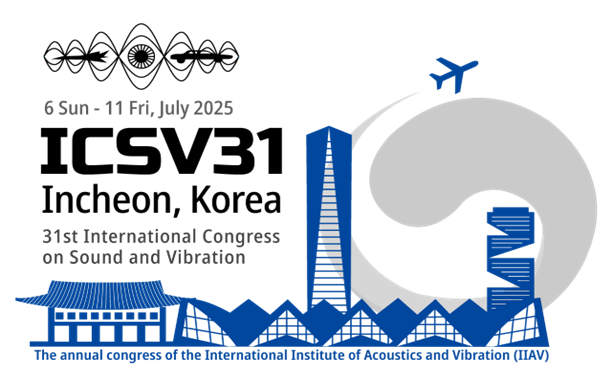}
    \end{flushright}
  \end{minipage}

  \vspace{6mm}

  {\fontsize{17pt}{20pt}\selectfont\sffamily \noindent \MakeUppercase{\textbf{\@title}}

  \vspace{3mm}\fontsize{14pt}{18pt}\selectfont\rmfamily \noindent \@author}
}
\makeatother

\titleformat{\section}
  {\fontsize{14}{14}\normalfont\sffamily\bfseries}
  {\thesection.}{1em}{}                

\titleformat{\subsection}
  {\normalfont\sffamily\bfseries}
  {\thesubsection}{1em}{}

\titleformat{\subsubsection}
  {\fontsize{12}{14}\selectfont\itshape}
  {\thesubsubsection}{1em}{}

\title{PERMUTATION-INVARIANT PHYSICS-INFORMED NEURAL NETWORK FOR REGION-TO-REGION SOUND FIELD RECONSTRUCTION}

\author{Xingyu CHEN, Sipei ZHAO, Fei MA, Eva CHENG and Ian S. BURNETT\\
{\small \textit{Centre for Audio, Acoustics and Vibration, University of Technology Sydney, Australia \\ email: Xingyu.Chen-8@student.uts.edu.au}}
}

\begin{document}

\maketitle
\renewcommand{\abstractname}{\vspace{-\baselineskip}} 

\begin{abstract}	\noindent
Most existing sound field reconstruction methods target point-to-region reconstruction, interpolating the Acoustic Transfer Functions (ATFs) between a fixed-position sound
source and a receiver region.
The applicability of these methods is limited because real-world ATFs tend to varying continuously with respect to the positions of sound sources and receiver regions. 
This paper presents a permutation-invariant physics-informed neural network for 
region-to-region sound field reconstruction, which aims to interpolate the 
ATFs across continuously varying sound sources and measurement regions.
The proposed method employs a deep set architecture to process the receiver and 
sound source positions as an unordered set, preserving acoustic reciprocity.
Furthermore, it incorporates the Helmholtz equation as a physical constraint to guide network training, ensuring physically consistent predictions. 
Experiments on real-world datasets demonstrate that with limited ATFs, the proposed 
method achieves more accurate reconstructions than the kernel-based method.
\vspace{1mm}


\end{abstract}

\vspace*{-5mm}
\quad\rule{425pt}{0.4pt}

\section{Introduction}
Sound Field Reconstruction (SFR), also known as sound field estimation or 
interpolation, aims to reconstruct the sound field over a spatial region within 
an acoustic environment~\cite{ueno2025sound}.
This task is the foundation for many applications, such as room acoustics~\cite{kuttruff2016room}, virtual and augmented reality systems~\cite{lentz2007virtual}, and speech dereverberation~\cite{habets2010speech}.
Under the assumptions of time-invariant acoustic properties and linear sound 
propagation, the frequency response of the Room Impulse Response (RIR) — 
known as the Acoustic Transfer Function (ATF) — effectively represents 
the sound field in the frequency domain.

Conventional SFR methods are mainly based on basis function decomposition. 
They decompose the sound field at some measurement positions onto basis 
functions, such as cylindrical harmonics~\cite{williams1999fourier}, 
spherical harmonics~\cite{ward2001reproduction}), 
plane waves~\cite{kirkeby1993reproduction}, 
and weighted kernel functions~\cite{caviedes2021gaussian}. 
These basis functions cover the entire receiver regions and thus allow the 
interpolation of the sound field at unmeasured positions. 
Novel methods, on the other hand, are mainly based on neural networks, 
especially the Physics-Informed Neural Networks (PINNs)~\cite{raissi2019physics}. 
They exploit the Helmholtz equation, the governing Partial Differential Equation of 
acoustic wave propagation in the loss function~\cite{koyama2024physics} to
regularize the training of neural networks and ensure physical consistency 
of the neural network output. 

However, most of the conventional and novel SFR methods are limited to 
the point-to-region scenario, where a fixed-position sound source is paired with 
varying receiver positions within a fixed measurement region~\cite{ueno2025sound}.
They have not effectively addressed the additional challenge introduced by the position variation of the sound sources and the measurement regions. 

The challenge has attracted researchers' attention and was tackled by several  studies~\cite{samarasinghe2015efficient,ribeiro2020kernel}. 
In \cite{samarasinghe2015efficient}, 
the ATFs between a sound source region and a receiver region were parameterized
as a truncated series of spherical harmonics.
The series does not depend on the positions of the sound source or the 
receiver, allowing the interpolation of ATFs between any positions 
within the sound and receiver regions.
However, separate analyses of the spherical harmonics truncation are required  
in each region, which can lead to error accumulation across regions. 
More recently, a Kernel Ridge Regression (KRR) approach~\cite{ribeiro2020kernel} 
was proposed to interpolate ATFs between regions using an infinite-dimensional representation. 
By directly estimating the reverberant field in a non-parametric manner, it avoids the empirical truncation of function series~\cite{samarasinghe2015efficient}. 

This paper presents a Permutation-Invariant PINN (PI-PINN) method, extending our previous SFR work~\cite{chen2023sound, ma2024sound} from the point-to-region scenario 
to region-to-region scenario. 
The proposed method employs a deep set architecture~\cite{zaheer2017deep} to 
process receiver and sound source positions as an unordered set, 
inherently preserving acoustic reciprocity. 
Additionally, the Helmholtz equation is incorporated as a physics constraint during training, ensuring physically consistent predictions. 
We evaluate the proposed PI-PINN method on real-world datasets and compare its performance with the current KRR method.


\section{Problem Formulation}

\begin{figure}[h]
\centering  
\includegraphics[width=6cm]{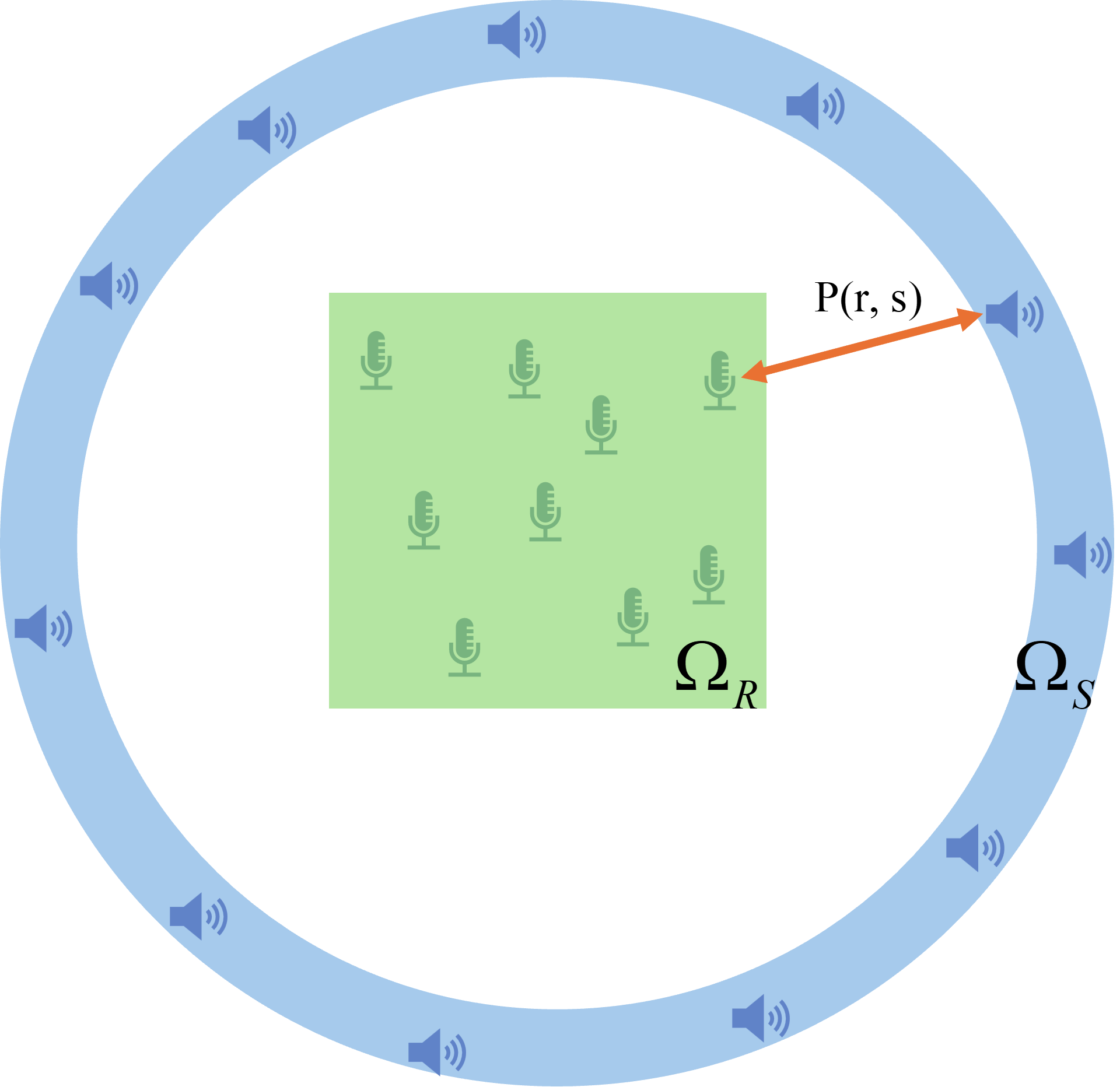}
\caption{Conceptual diagram of Region-to-Region SFR.}
\label{fig:concept}
\end{figure}

As illustrated in Fig.~\ref{fig:concept}, region-to-region SFR aims to reconstruct the ATFs from an arbitrary sound source in the source domain ($\Omega_{S}$) to an arbitrary receiver in the receiver domain ($\Omega_{R}$) based on limited measurements.
We denote the ATF by $P(\mathbf{r}, \mathbf{s}, f)$, where $\mathbf{r} \in \Omega_{R}$ represents a receiver position, $\mathbf{s} \in \Omega_{S}$ represents a source position, and $f$ is the frequency. 
In practice, the ATF is only measured at a finite set of receiver–source pairs \(\{(\mathbf{r}_i, \mathbf{s}_i)\}_{i=1}^{N}\).

The ATF \(P(\mathbf{r}, \mathbf{s}, f)\) is governed by the Helmholtz equation, which describes wave behavior in a linear, isotropic, and lossless medium:
\begin{equation}
\nabla^2 P(\mathbf{r}, \mathbf{s}, f) + k^2 P(\mathbf{r}, \mathbf{s}, f) = 0,
\label{eq:helmholtz}
\end{equation}
where \(\nabla^2\) denotes the Laplacian operator with respect to spatial coordinates, \(k = \frac{2\pi f}{c}\) is the wavenumber, and \(c\) is the speed of sound.

For a point sound source, the ATF satisfies the reciprocity principle, which states that the transfer function is symmetric with respect to the source and receiver positions:
\begin{equation}
P(\mathbf{r}, \mathbf{s}, f) = P(\mathbf{s}, \mathbf{r}, f).
\label{eq:reciprocity}
\end{equation}

The objective of region-to-region SFR is to predict $\hat{P}(\mathbf{r}, \mathbf{s}, f)$ for arbitrary source and receiver locations across the entire domains, i.e.,  $\forall(\textbf{r}, \textbf{s}) \in \Omega_{R} \times \Omega_{S}$, using limited measurements $P(\mathbf{r}_i, \mathbf{s}_i, f_i)$ at \(\{(\mathbf{r}_i, \mathbf{s}_i)\}_{i=1}^{N}\). Specifically, the predictions should align with the measured data $P(\mathbf{r}_i, \mathbf{s}_i, f_i)$,  satisfy the Helmholtz equation in Eq.~\eqref{eq:helmholtz}, and preserve the symmetry constraint in Eq.~\eqref{eq:reciprocity}.

\section{Methodology}

\begin{figure}[h]
\centering  
\includegraphics[width=12cm]{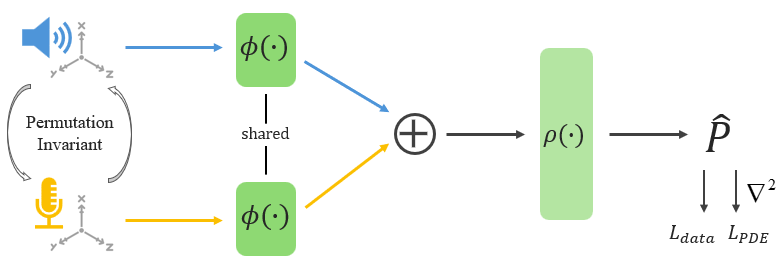}
\caption{Conceptual diagram of the proposed PI-PINN. }
\label{fig2}
\end{figure}

We propose a PI-PINN for region-to-region SFR, as illustrated in Fig.~\ref{fig2}. 
The method combines a deep set architecture and a physics loss function. 
The deep set architecture ensures permutation invariance by processing 
receiver and sound source positions as an unordered set to satisfy the 
acoustic reciprocity.
The physics loss function incorporates the Helmholtz equation as a 
regularization term to ensure physically consistent predictions. 
Further details are provided in the subsequent subsections.

\subsection{Network Architecture for Permutation Invariance}
The proposed method employs a deep set architecture~\cite{zaheer2017deep} to 
achieve permutation invariance in processing receiver and sound source positions. 
Deep sets provide a framework for modeling functions over unordered sets, 
ensuring that the output remains invariant to the permutation of input elements. 
This architecture has been successfully applied in various domains, such as point 
cloud processing~\cite{guo2020deep} and array processing~\cite{luo2020end}.



We treat the receiver and sound source positions 
as elements of an unordered set, $\{\textbf{r}, \textbf{s}\}$, enabling the 
capture of the spatial acoustic relationship without relying on fixed ordering. 
By design, the deep set architecture respects the symmetry of the ATF under the 
exchange of receiver and sound source positions, a direct result of the acoustic reciprocity principle.

As illustrated in Fig.~\ref{fig2}, the method consists of two sub-networks, $\phi(\cdot)$ and $\rho(\cdot)$, 
both parameterized by multi-layer perceptrons (MLPs) with the $\tanh$ activation function.
The sub-network $\phi$ is a feature extraction function that maps each input 
coordinate to a latent feature space, 
while the main network $\rho$ aggregates the resulting features through summation:
\begin{equation}
\hat{P}(\mathbf{r}, \mathbf{s}, f) = \rho\left(\phi(\mathbf{r}) + \phi(\mathbf{s})\right).
\label{eq:deepsets}
\end{equation}
The summation operation ensures permutation invariance, 
such that exchanging the receiver and sound source position, $\mathbf{r}$ and $\mathbf{s}$, does not alter the feature representation. 
This property inherently respects the acoustic reciprocity principle without requiring explicit constraints in the loss function. The output $\hat{P}(\mathbf{r}, \mathbf{s}, f)$ is the predicted ATF value for the given receiver and source positions.

\subsection{Physics-Informed Neural Network}

We use a neural network to learn the mapping from $(\mathbf{r}, \mathbf{s}, f)$ to the sound pressure $P(\mathbf{r}, \mathbf{s}, f)$.
To ensure physically consistent predictions, the method incorporates the Helmholtz equation as a physical loss in addition to the data loss so that the main network $\rho(\cdot)$ is a PINN. PINNs extend the capabilities of neural fields by embedding physical laws into the training process through partial differential equations (PDEs) as regularization terms. 
By leveraging automatic differentiation, the PDE loss is evaluated based on the input coordinates, guiding the network's outputs to align with the governing physical equations.

In the proposed method, we incorporate the Helmholtz equation in Eq.\eqref{eq:reciprocity} as a physical loss, which is defined as the mean squared residual (MSE) of:
\begin{equation}
L_{\text{PDE}} = \frac{1}{N_{\text{PDE}}} \sum_{i=1}^{N_{\text{PDE}}} \biggl\| \nabla^2 \hat{P}(\mathbf{r}_i, \mathbf{s}_i, f_i) + \left( \frac{2\pi f_i}{c} \right)^2 \hat{P}(\mathbf{r}_i, \mathbf{s}_i, f_i) \biggr\|^2,
\label{eq:pde_loss}
\end{equation}
where $N_{\text{PDE}}$ denotes the number of sampled points that are uniformly and densely sampled within the interpolation domain.

In addition to $L_{\text{PDE}}$, a data loss is introduced to ensure that the network predictions, $\hat{h}_\theta(\mathbf{r}_i, \mathbf{s}_i, f_i)$, align with the measured ATF values $P(\mathbf{r}_i, \mathbf{s}_i, f_i)$ at the sampled receiver-source-frequency triplets $\{(\mathbf{r}_i, \mathbf{s}_i, f_i)\}_{i=1}^{N_{\text{data}}}$. The data loss is defined as:
\begin{equation}
L_{\text{data}} = \frac{1}{N_{\text{data}}} \sum_{i=1}^{N_{\text{data}}} \bigl\|\hat{P}_\theta(\mathbf{r}_i, \mathbf{s}_i, f_i) - P(\mathbf{r}_i, \mathbf{s}_i, f_i)\bigr\|^2,
\label{eq:data_loss}
\end{equation}
where $N_{\text{data}}$ denotes the number of measured ATFs. This term encourages the network to produce predictions to match the measured ATF values, ensuring consistency with the observed data.

The total loss for training the network is a weighted combination of the data loss and the physics-informed loss:
\begin{equation}
L_{\text{total}} = L_{\text{data}} + \lambda L_{\text{PDE}},
\label{eq:total_loss}
\end{equation}
where $\lambda$ is a hyperparameter that balances the contribution of the physics-informed regularization. 
Based on our previous work~\cite{ma2024sound}, we set $\lambda$ to 1 for experiments.
By minimizing $L_{\text{total}}$, the network learns to satisfy both the observed data and the underlying physics, resulting in accurate and physically consistent ATF predictions.

\subsection{Implementation Remarks}

The proposed method is implemented using TensorFlow and leverages nested gradient tapes to compute the second-order derivatives required for the physical loss in Eq.~\eqref{eq:pde_loss}. Our method comprises two sub-networks, $\phi$ and $\rho$, each implemented as a multi-layer perceptron (MLP) with two hidden layers of 128 neurons. We chose $\tanh(\cdot)$ activation functions throughout because they support the computation of the second-order gradients needed for $L_{\text{PDE}}$. Although we experimented with $\sin(\cdot)$ activation functions~\cite{sitzmann2020implicit}—commonly used in neural field methods~\cite{di2024neural}—their incorporation with $L_{\text{PDE}}$ resulted in unstable training dynamics and convergence issues.

For the output, we adopt a format that outputs either the real or the imaginary part of the sound pressure for a single frequency bin, necessitating the training of $F \times 2$ separate models (with $F$ representing the number of frequency bins). Our network design and hyperparameter choices were guided by preliminary experiments; further tuning and the exploration of alternative architectures will be pursued in future work.

\section{Experiments}
In this section, we compare the performance of the proposed PI-PINN method with the KRR method using the University of Technology Sydney (UTS) multi-zone sound field reproduction dataset~\cite{zhao2022room}. 
First, we conduct an ablation study to assess the contribution of key components in our method. 
Second, we compare the proposed method against a baseline KRR interpolation method~\cite{ribeiro2020kernel}.

\subsection{Dataset}

\begin{figure}[h]
    \centering
    \begin{minipage}[t]{0.48\textwidth}
    \centering
    \includegraphics[width=6cm,height=4cm]{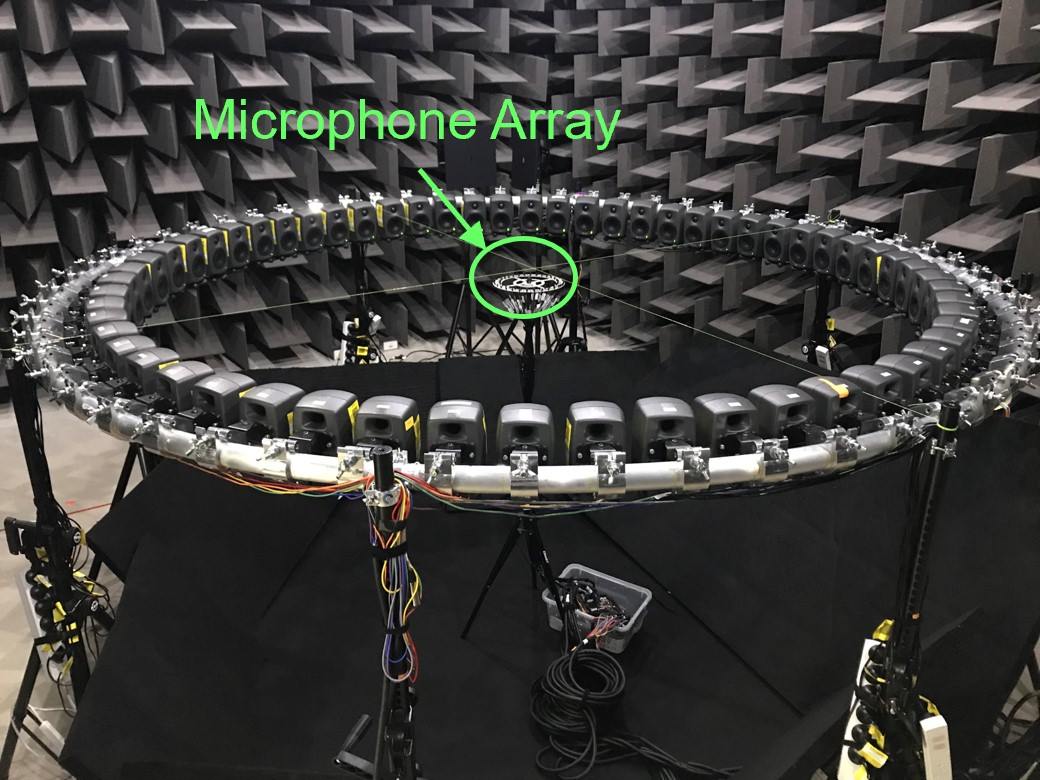}
    \\\small (a) 
\end{minipage}
\hfill
\begin{minipage}[t]{0.48\textwidth}
    \centering
    \includegraphics[width=6cm,height=4cm]{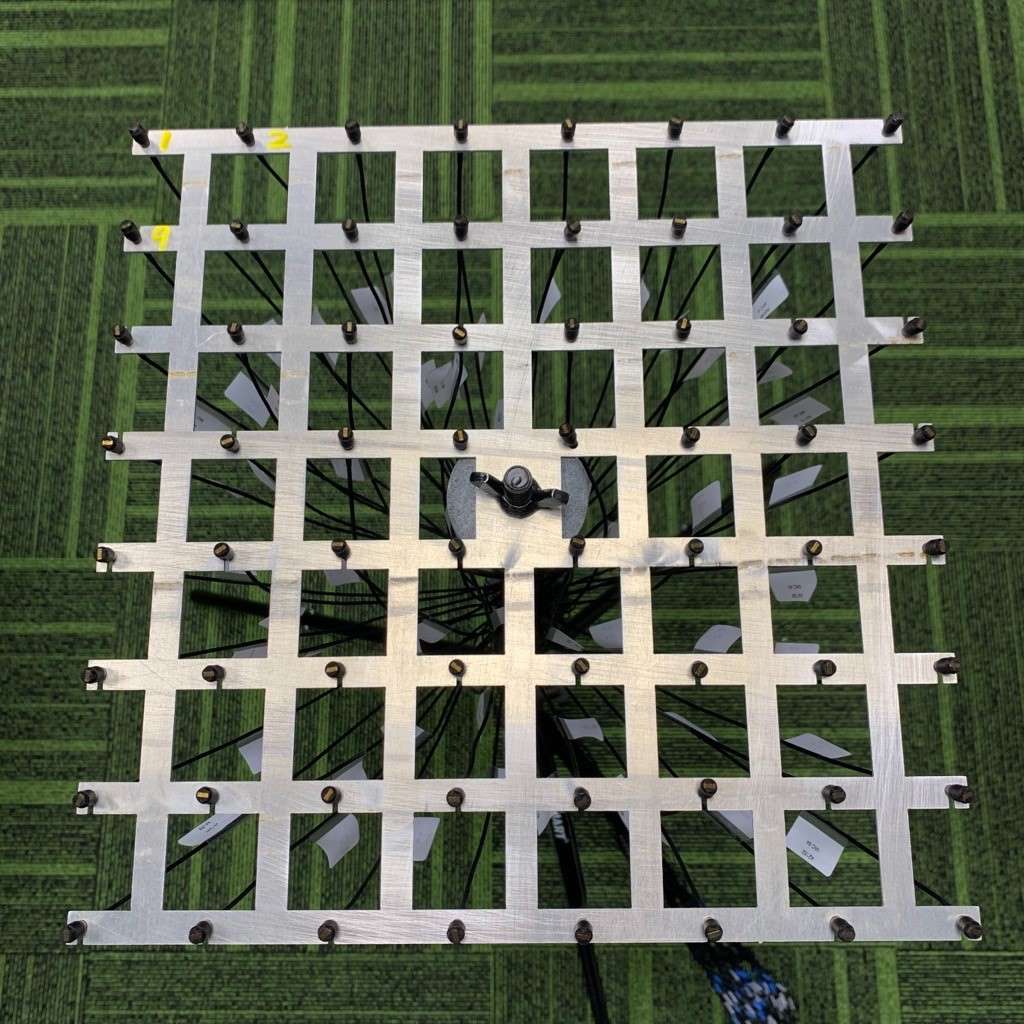}
    \\\small (b) 
\end{minipage}

\caption{(a) Measurement setup from the UTS dataset,
(b) Planar 64-microphone array.}

\label{fig:setup}
    \vspace{-0.5em}
\end{figure}

Fig.~\ref{fig:setup} illustrates the experimental setup of the UTS dataset.  
We use the RIRs recorded between a circular $60$-loudspeaker array and a planar $64$-microphone array.  
The loudspeaker array is uniformly arranged on a circle with a radius of $1.5~\mathrm{m}$.  
The microphone array, with its center located at the origin (corresponding to Zone E as defined in the dataset ~\cite{zhao2022room}), spans a square with a side length of $0.28~\mathrm{m}$, and the microphones are uniformly distributed in a grid pattern with an interval of $0.04~\mathrm{m}$.

We transformed the first 0.5 s of each RIR into the frequency domain and 
retained the positive frequency components.
The resulting complex-valued ATFs are divided into two groups: the training set contains ATFs from $30$ evenly spaced 
loudspeakers to the 28 edge microphones, while the testing set consists of the ATFs from another $30$ loudspeakers to all the 64 microphones.


The Normalized MSE (NMSE) is used as the evaluation metric, which is defined as:
\begin{equation}
\text{NMSE}(f) = 10 \log_{10} \frac{\sum_{n=1}^{N} \|P(q_n, f) - \hat{P}(q_n, f)\|^2}{\sum_{n=1}^{N} \|P(q_n, f)\|^2},
\label{eq:nmse}
\end{equation}
where $N = 30\times 64 = 1920$ is the total number of ATFs in the testing set, $q_n$ represents the $n$-th test source/receiver pair, $P(q_n, f)$ is the ground truth transfer function, and $\hat{P}(q_n, f)$ is the corresponding predicted results.

\subsection{Ablation Study}

\begin{figure*}[h]
\begin{center}
\includegraphics[width=16cm]{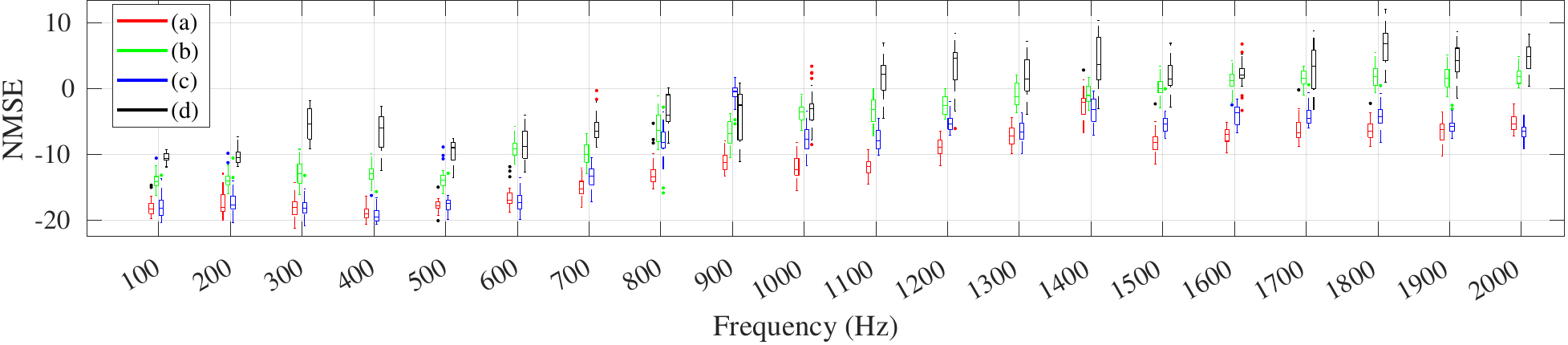}
\end{center}
\vspace{-1.5em}
\caption{NMSE (dB) as a function of frequency for four model variants in an anechoic room.}
\label{fig:Ablation}
\end{figure*}

To assess the impact of key components in our proposed PI-PINN method shown in Fig.~\ref{fig:Ablation}, we conduct an ablation study by evaluating the performance of the following four model variants in an anechoic room:
\begin{itemize}
    \item[(a)] The full PI-PINN model (Fig.~\ref{fig:Ablation}) with both the permutation invariant network \( \phi(\cdot) \) and \( L_{\text{PDE}} \).
    \item[(b)] The permutation invariant network with \( \phi(\cdot) \), but \textbf{without} \( L_{\text{PDE}} \) regularization.
    \item[(c)] The PINN model with \( L_{\text{PDE}} \), but \textbf{without} the permutation invariant network \( \phi(\cdot) \).
    \item[(d)] A plain feedforward network, \textbf{without} the permutation invariant network \( \phi(\cdot) \) or \( L_{\text{PDE}} \).
\end{itemize}


The NMSE results of the four model variants are presented in Fig.~\ref{fig:Ablation}. 
The proposed PI-PINN model (a) achieves the best performance, demonstrating the importance of both physical regularization and reciprocity.
Removing the physical loss \(L_{\text{PDE}}\) (model b) causes a significant performance drop, suggesting that the model overfits the training data and fails to generalize in unmeasured regions. 
Meanwhile, the performance gap between (a) and (c), as well as between (b) and (d), illustrates the contribution of the permutation-invariant architecture in enforcing acoustic reciprocity.


\vspace{-0.5em}
\subsection{Results}
\vspace{-0.5em}

\begin{figure*}[h]
    \centering

    \begin{minipage}{0.95\textwidth}
        \centering
        \includegraphics[width=16cm]{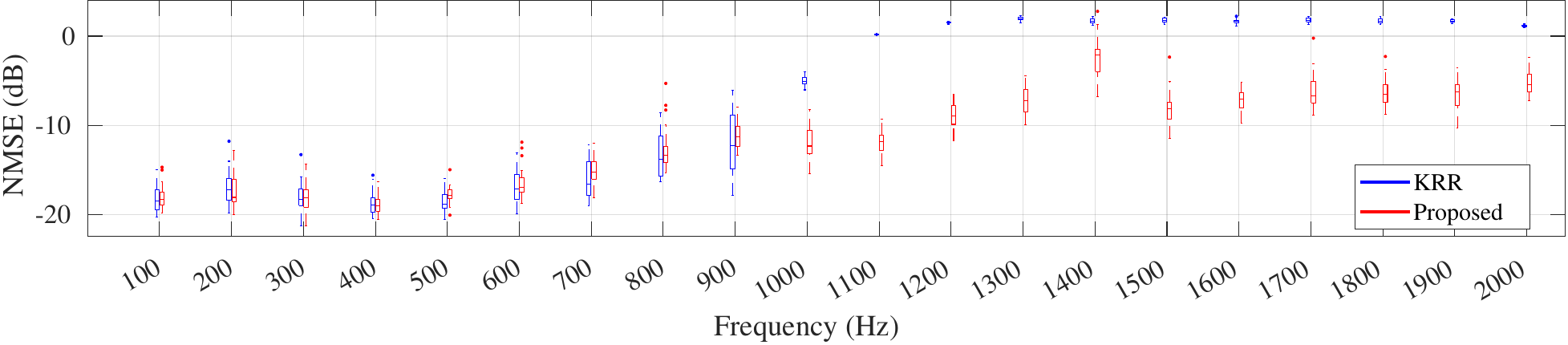}
        \\\small (a) Anechoic room. 
    \end{minipage}


    \begin{minipage}{0.95\textwidth}
        \centering
        \includegraphics[width=16cm]{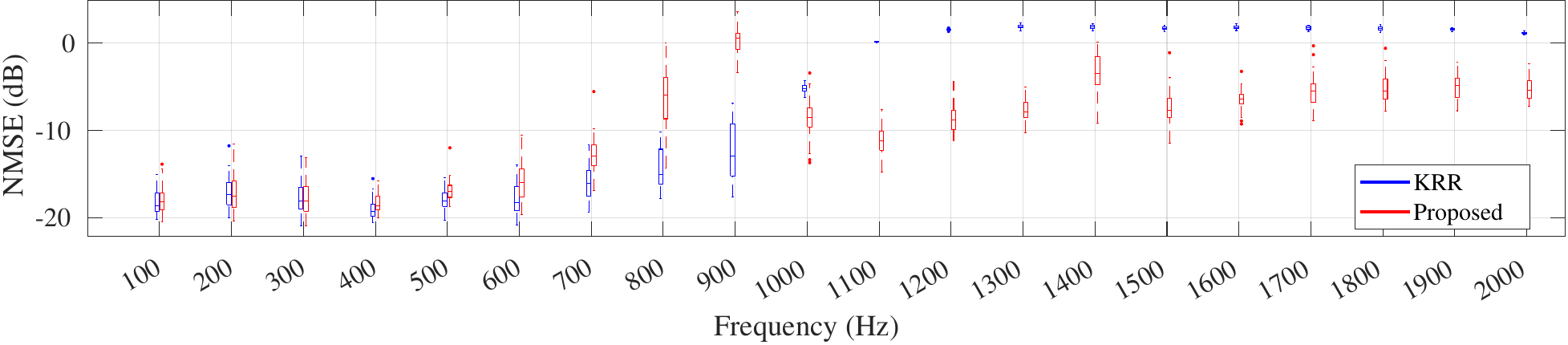}
        \\\small (b) Hemi-anechoic room.
    \end{minipage}

    \caption{NMSE (dB) as a function of frequency for the proposed PI-PINN model and the KRR method.}
    \label{fig:5}
\end{figure*}

We compare the proposed PI-PINN model against the KRR method~\cite{ribeiro2022region}. Figure~\ref{fig:5}(a) and~\ref{fig:5}(b) present the NMSE (dB) against frequency for the two methods in the anechoic (top) and hemi-anechoic (bottom) rooms, respectively. Fig.~\ref{fig:5}(a) shows that, in the anechoic room condition, the proposed PI-PINN exhibits comparable performance with the KRR method below approximately $1000$~Hz but begins to outperform at frequencies above $1100$~Hz. The NMSE above $1000$~Hz is higher than 0 dB for the KRR method, showing its failure in reconstructing the sound field. By contrast, the proposed PI-PINN method improves the average NMSE by more than 5 dB at most frequencies except 1400 Hz. Similar results can be observed in Fig.~\ref{fig:5}(b) for the hemi-anechoic chamber except at 800 Hz and 900 Hz, where the proposed PI-PINN model performs worse than the KRR method. The reason for this inferior performance at this specific frequency range is unclear yet and will be investigated in our future work.

\begin{figure}[htbp]
    \centering
    \begin{minipage}[t]{0.32\textwidth}
        \centering
        \includegraphics[width=5.5cm]{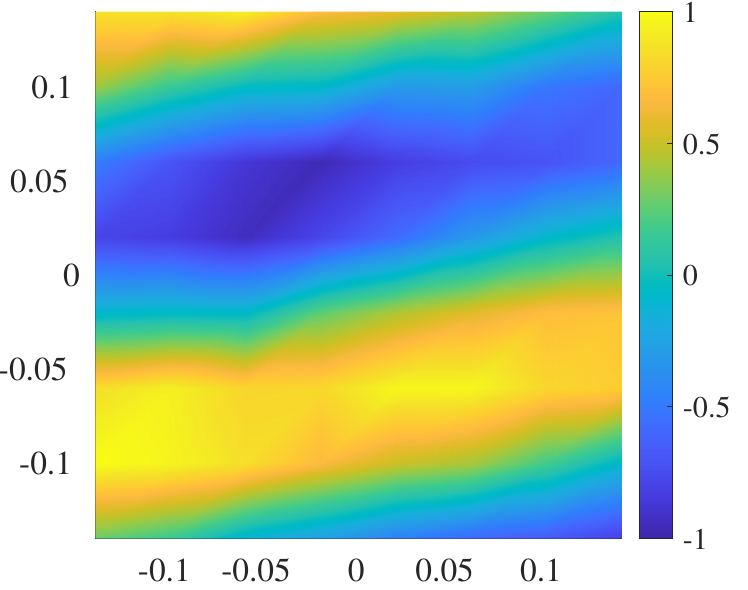}
        \\ \small (a) Anechoic room - measured
    \end{minipage}
    \hfill
    \begin{minipage}[t]{0.32\textwidth}
        \centering
        \includegraphics[width=5.5cm]{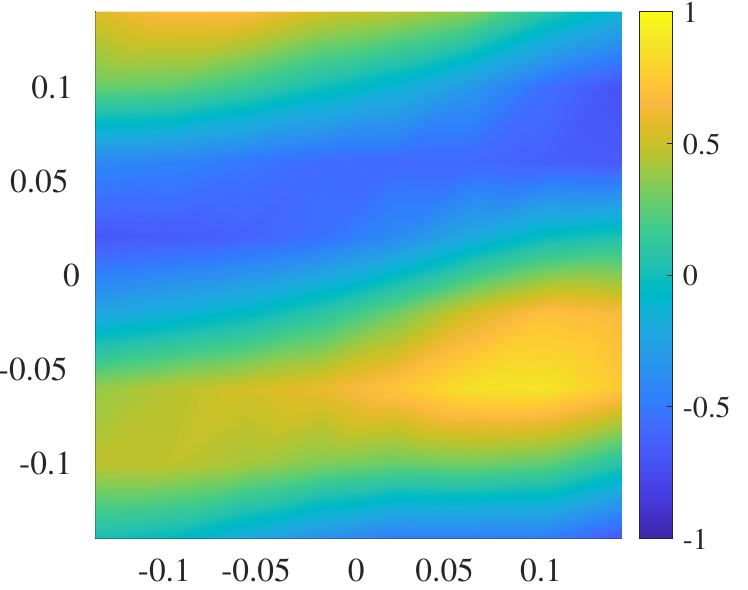}
        \\ \small (b)  Anechoic room - Proposed
    \end{minipage}
    \hfill
    \begin{minipage}[t]{0.32\textwidth}
        \centering
        \includegraphics[width=5.5cm]{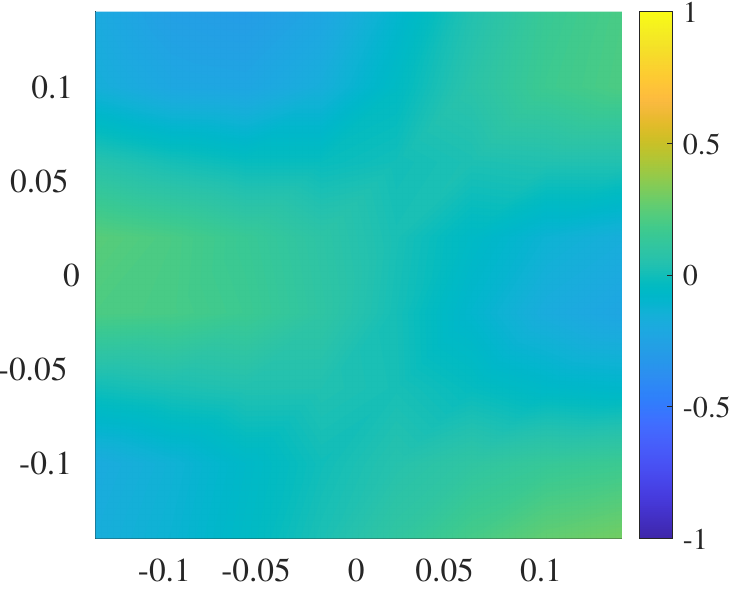}
        \\ \small (c)  Anechoic room - KRR
    \end{minipage}
    
    \vspace{0.5em} 
    
    \begin{minipage}[t]{0.32\textwidth}
        \centering
        \includegraphics[width=5.5cm]{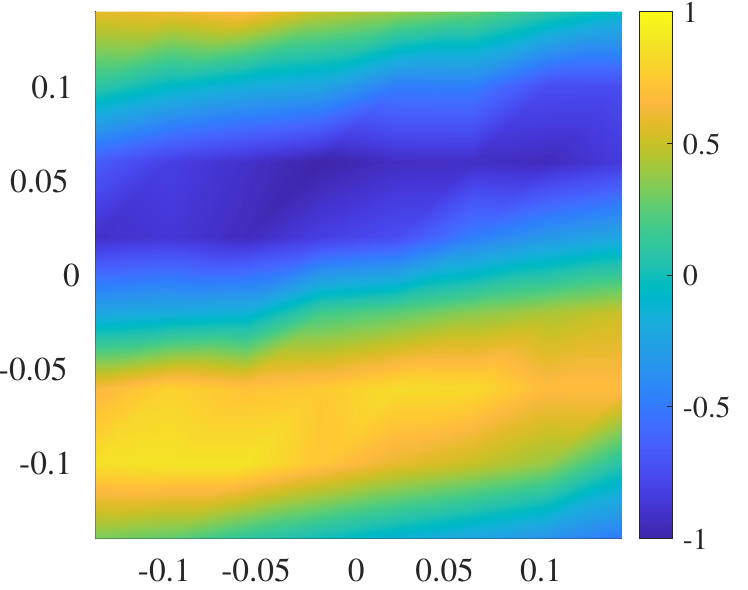}
        \\ \small (d) Hemi-anechoic room - measured
    \end{minipage}
    \hfill
    \begin{minipage}[t]{0.32\textwidth}
        \centering
        \includegraphics[width=5.5cm]{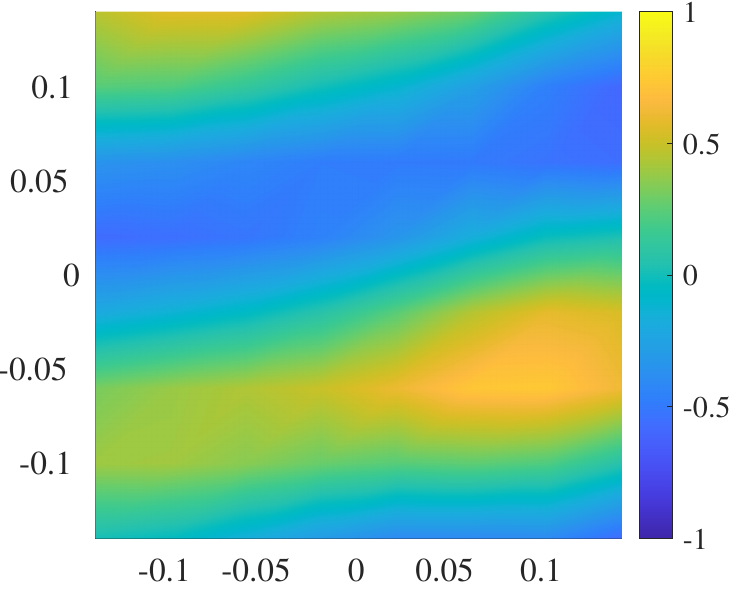}
        \\ \small (e)  Hemi-anechoic room - Proposed
    \end{minipage}
    \hfill
    \begin{minipage}[t]{0.32\textwidth}
        \centering
        \includegraphics[width=5.5cm]{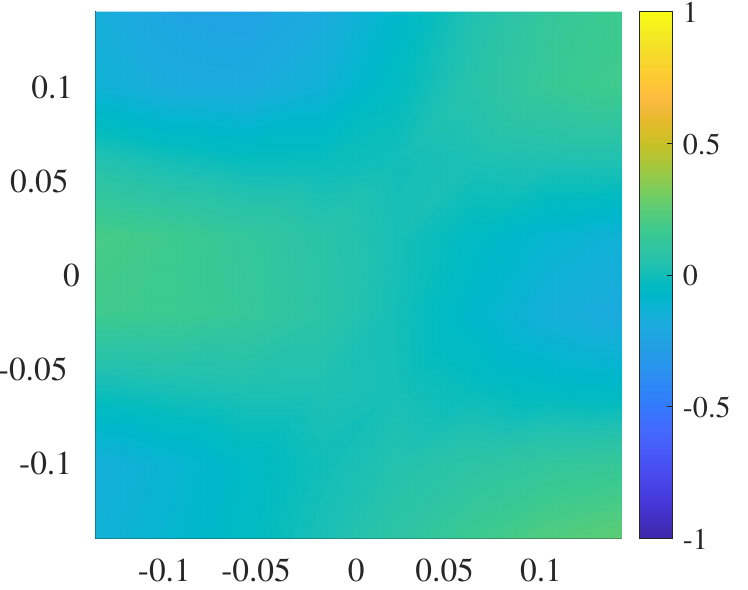}
        \\ \small (f)  Hemi-anechoic room - KRR
    \end{minipage}

    \caption{Measured and Reconstructed ATF distribution (real part) for Loudspeaker $\#2$ at $1500$~Hz.}
    \label{fig:atf_reconstruction}
\end{figure}

Figure~\ref{fig:atf_reconstruction} shows the reconstructed ATF distribution for Loudspeaker $\#2$ at $1500$~Hz. 
The measured ground truth distributions (a) and (d) exhibit clear wave-like structures with distinct spatial layers, including two peaks. 
The reconstructions produced by the proposed method (b) and (e) align well with the ground truth in both the overall shape and position of key features, such as the blue low-pressure regions. Some minor smoothing and blurring are observed near edges, but the dominant spatial patterns are well preserved. 
In contrast, the KRR results (c) and (f) fail to capture the main structural features, where the reconstructed fields are overly smooth and generally close to zero, indicating a loss of spatial variation and an underestimation of ATFs.

\vspace{-1 em}



\vspace{-1 em}
\section{Conclusion}
\vspace{-0.5 em}

This paper proposes a PI-PINN method for region-to-region SFR. 
By incorporating a permutation-invariant architecture, 
the proposed method respects acoustic reciprocity while using the Helmholtz equation to enforce physical consistency.
Experimental results demonstrate the superiority of our approach over the existing KRR method, particularly in the higher frequency range between $1000$~Hz and $2000$~Hz.
Compared to the baseline KRR method, our method better captures spatial variations of sound fields.

Although improvements were observed in controlled acoustic environments (i.e., anechoic and hemi-anechoic rooms), the performance in complex room settings (e.g., meeting rooms) remains limited for both the KRR and proposed methods. 
In future work, we plan to extend our framework to realistic indoor environments. 
These realistic environments pose additional challenges due to complex geometries and furniture, which introduce extra reflections and scattering. 
To address these issues, future work will incorporate environment-specific priors and develop frequency-dependent adaptations.

\vspace{-0.5em}
\section{Acknowledgments}
\vspace{-0.5em}

Computational facilities were provided by the UTS eResearch High Performance Computer Cluster.

\clearpage

\makeatletter
\let\oldthebibliography\thebibliography
\renewcommand{\thebibliography}[1]{%
  \oldthebibliography{#1}%
  \setlength{\itemsep}{6pt}     
  \setlength{\parskip}{0pt}     
  \setlength{\parsep}{0pt}      
}
\makeatother

\vspace*{-3em}
{
\fontsize{10}{12}\selectfont
\bibliographystyle{icsv_bib}
\bibliography{references}

@article{zaheer2017deep,
  title={Deep sets},
  author={Zaheer, Manzil and Kottur, Satwik and Ravanbakhsh, Siamak and Poczos, Barnabas and Salakhutdinov, Russ R and Smola, Alexander J},
  journal={Advances in neural information processing systems},
  volume={30},
  year={2017}
}

@article{zhao2022room,
  title={A room impulse response database for multizone sound field reproduction (L)},
  author={Zhao, Sipei and Zhu, Qiaoxi and Cheng, Eva and Burnett, Ian S},
  journal={The Journal of the Acoustical Society of America},
  volume={152},
  number={4},
  pages={2505--2512},
  year={2022},
  publisher={AIP Publishing}
}

@inproceedings{chen2023sound,
  title={Sound field estimation around a rigid sphere with physics-informed neural network},
  author={Chen, Xingyu and Ma, Fei and Bastine, Amy and Samarasinghe, Prasanga and Sun, Huiyuan},
  booktitle={2023 Asia Pacific Signal and Information Processing Association Annual Summit and Conference (APSIPA ASC)},
  pages={1984--1989},
  year={2023},
  organization={IEEE}
}

@article{ma2024sound,
  title={Sound field reconstruction using a compact acoustics-informed neural network},
  author={Ma, Fei and Zhao, Sipei and Burnett, Ian S},
  journal={The Journal of the Acoustical Society of America},
  volume={156},
  number={3},
  pages={2009--2021},
  year={2024},
  publisher={AIP Publishing}
}

@article{sitzmann2020implicit,
  title={Implicit neural representations with periodic activation functions},
  author={Sitzmann, Vincent and Martel, Julien and Bergman, Alexander and Lindell, David and Wetzstein, Gordon},
  journal={Advances in neural information processing systems},
  volume={33},
  pages={7462--7473},
  year={2020}
}

@inproceedings{di2024neural,
  title={NEURAL STEERER: NOVEL STEERING VECTOR SYNTHESIS WITH A CAUSAL NEURAL FIELD OVER FREQUENCY AND DIRECTION},
  author={Di Carlo, Diego and Nugraha, Aditya Arie and Fontaine, Mathieu and Bando, Yoshiaki and Yoshii, Kazuyoshi},
  booktitle={ICASSP},
  year={2024}
}

@article{ribeiro2022region,
  title={Region-to-region kernel interpolation of acoustic transfer functions constrained by physical properties},
  author={Ribeiro, Juliano GC and Ueno, Natsuki and Koyama, Shoichi and Saruwatari, Hiroshi},
  journal={IEEE/ACM Transactions on Audio, Speech, and Language Processing},
  volume={30},
  pages={2944--2954},
  year={2022},
  publisher={IEEE}
}

@inproceedings{ribeiro2020kernel,
  title={Kernel interpolation of acoustic transfer function between regions considering reciprocity},
  author={Ribeiro, Juliano GC and Ueno, Natsuki and Koyama, Shoichi and Saruwatari, Hiroshi},
  booktitle={2020 IEEE 11th Sensor Array and Multichannel Signal Processing Workshop (SAM)},
  pages={1--5},
  year={2020},
  organization={IEEE}
}

@article{ueno2025sound,
  title={Sound Field Estimation: Theories and Applications},
  author={Ueno, Natsuki and Koyama, Shoichi and others},
  journal={Foundations and Trends{\textregistered} in Signal Processing},
  volume={19},
  number={1},
  pages={1--98},
  year={2025},
  publisher={Now Publishers, Inc.}
}

@article{caviedes2021gaussian,
  title={Gaussian processes for sound field reconstruction},
  author={Caviedes-Nozal, Diego and Riis, Nicolai AB and Heuchel, Franz M and Brunskog, Jonas and Gerstoft, Peter and Fernandez-Grande, Efren},
  journal={The Journal of the Acoustical Society of America},
  volume={149},
  number={2},
  pages={1107--1119},
  year={2021},
  publisher={AIP Publishing}
}

@article{koyama2024physics,
  title={Physics-informed machine learning for sound field estimation},
  author={Koyama, Shoichi and Ribeiro, Juliano GC and Nakamura, Tomohiko and Ueno, Natsuki and Pezzoli, Mirco},
  journal={arXiv preprint arXiv:2408.14731},
  year={2024}
}

@article{raissi2019physics,
  title={Physics-informed neural networks: A deep learning framework for solving forward and inverse problems involving nonlinear partial differential equations},
  author={Raissi, Maziar and Perdikaris, Paris and Karniadakis, George E},
  journal={Journal of Computational physics},
  volume={378},
  pages={686--707},
  year={2019},
  publisher={Elsevier}
}

@book{williams1999fourier,
  title={Fourier acoustics: sound radiation and nearfield acoustical holography},
  author={Williams, Earl G},
  year={1999},
  publisher={Elsevier}
}

@article{samarasinghe2015efficient,
  title={An efficient parameterization of the room transfer function},
  author={Samarasinghe, Prasanga and Abhayapala, Thushara and Poletti, Mark and Betlehem, Terence},
  journal={IEEE/ACM Transactions on Audio, Speech, and Language Processing},
  volume={23},
  number={12},
  pages={2217--2227},
  year={2015},
  publisher={IEEE}
}

@incollection{habets2010speech,
  title={Speech dereverberation using statistical reverberation models},
  author={Habets, Emanu{\"e}l AP},
  booktitle={Speech dereverberation},
  pages={57--93},
  year={2010},
  publisher={Springer}
}

@book{kuttruff2016room,
  title={Room acoustics},
  author={Kuttruff, Heinrich},
  year={2016},
  publisher={Crc Press}
}

@article{lentz2007virtual,
  title={Virtual reality system with integrated sound field simulation and reproduction},
  author={Lentz, Tobias and Schr{\"o}der, Dirk and Vorl{\"a}nder, Michael and Assenmacher, Ingo},
  journal={EURASIP journal on advances in signal processing},
  volume={2007},
  pages={1--19},
  year={2007},
  publisher={Springer}
}

@article{ward2001reproduction,
  title={Reproduction of a plane-wave sound field using an array of loudspeakers},
  author={Ward, Darren B and Abhayapala, Thushara D},
  journal={IEEE Transactions on speech and audio processing},
  volume={9},
  number={6},
  pages={697--707},
  year={2001},
  publisher={IEEE}
}

@article{kirkeby1993reproduction,
  title={Reproduction of plane wave sound fields},
  author={Kirkeby, Ole and Nelson, Philip A},
  journal={The Journal of the Acoustical Society of America},
  volume={94},
  number={5},
  pages={2992--3000},
  year={1993},
  publisher={Acoustical Society of America}
}

@article{guo2020deep,
  title={Deep learning for 3d point clouds: A survey},
  author={Guo, Yulan and Wang, Hanyun and Hu, Qingyong and Liu, Hao and Liu, Li and Bennamoun, Mohammed},
  journal={IEEE transactions on pattern analysis and machine intelligence},
  volume={43},
  number={12},
  pages={4338--4364},
  year={2020},
  publisher={IEEE}
}

@inproceedings{luo2020end,
  title={End-to-end microphone permutation and number invariant multi-channel speech separation},
  author={Luo, Yi and Chen, Zhuo and Mesgarani, Nima and Yoshioka, Takuya},
  booktitle={ICASSP 2020-2020 IEEE international conference on acoustics, speech and signal processing (ICASSP)},
  pages={6394--6398},
  year={2020},
  organization={IEEE}
}
}

\label{LastPage}

\end{document}